\documentclass[%
 aip,
 amsmath,amssymb,
 reprint,%
]{revtex4-2}

\usepackage{graphicx}
\usepackage{dcolumn}
\usepackage{bm}

\usepackage[utf8]{inputenc}
\usepackage[T1]{fontenc}
\usepackage{mathptmx}
\usepackage{etoolbox}
\usepackage{siunitx}
\usepackage{microtype}
\usepackage{xcolor}
\usepackage[linesnumbered, ruled]{algorithm2e}
\SetKwRepeat{Do}{do}{while}

\makeatletter
\def\@email#1#2{%
 \endgroup
 \patchcmd{\titleblock@produce}
  {\frontmatter@RRAPformat}
  {\frontmatter@RRAPformat{\produce@RRAP{*#1\href{mailto:#2}{#2}}}\frontmatter@RRAPformat}
  {}{}
}%
\makeatother

\definecolor{purp}{rgb}{0.4,0.2,0.8}

\newcommand{\nelec}{n_\text{e}}
\newcommand{\nprot}{n_\text{p}}

\begin{document}

\preprint{AIP/123-QED}

\title[Less can be more: Insights on the role of electrode microstructure in redox flow batteries from 2D direct numerical simulations]{Less can be more: Insights on the role of electrode microstructure in redox flow batteries from 2D direct numerical simulations}

\author{Simone Dussi}
\affiliation{ 
John A. Paulson School of Engineering and Applied Sciences, Harvard University, 29 Oxford Street, Cambridge, MA 02138, United States
}%
\author{Chris H. Rycroft}%
\email{chr@seas.harvard.edu}
\affiliation{ 
John A. Paulson School of Engineering and Applied Sciences, Harvard University, 29 Oxford Street, Cambridge, MA 02138, United States
}%
\affiliation{ 
Mathematics Group, Lawrence Berkeley National Laboratory, 1 Cyclotron Road, Berkeley, CA 94720, United States
}
\date{\today}

\begin{abstract}
  Understanding how to structure a porous electrode to facilitate fluid, mass, and charge transport is key to enhance the performance of electrochemical devices such as fuel cells, electrolyzers, and redox flow batteries (RFBs). Using a parallel computational framework, direct numerical simulations are carried out on idealized porous electrode microstructures for RFBs. Strategies to improve electrode design starting from a regular lattice are explored. We observe that by introducing vacancies in the ordered arrangement, it is possible to achieve higher voltage efficiency at a given current density, thanks to improved mixing of reactive species, despite reducing the total reactive surface. Careful engineering of the location of vacancies, resulting in a density gradient, outperforms disordered configurations. Our simulation framework is a new tool to explore transport phenomena in RFBs and our findings suggest new ways to design performant electrodes.
\end{abstract}

\maketitle


\section{\label{sec:intro}Introduction}
In a more sustainable energy future, electricity produced by renewable sources, such as solar or wind, is expected to be widely spread and to permeate the entire electrical grid. However, to overcome the issues associated to the variability of these natural sources, a suitable (i.e.\@ reliable, efficient, safe, and inexpensive) energy storage technology is required. Redox flow batteries (RFBs) hold a great promise to reach such requirements.~\cite{Weber2011, Soloveichik2015, Sanchez-Diez2021} RFBs have a flexible design that allows power and energy capacities to be independently tuned. They can have an extremely long life-time, and can be made from chemicals that are non-toxic, non-flammable and composed of earth-abundant elements, making them promising candidates for competitive energy storage technology.~\cite{Huskinson2014, Lin2015, Janoschka2015, Kwabi2020} Yet, to completely fulfill their promise, RFBs need further research to improve their performance, e.g.\@ to increase their power density.

\begin{figure}
\includegraphics[width=0.4\textwidth]{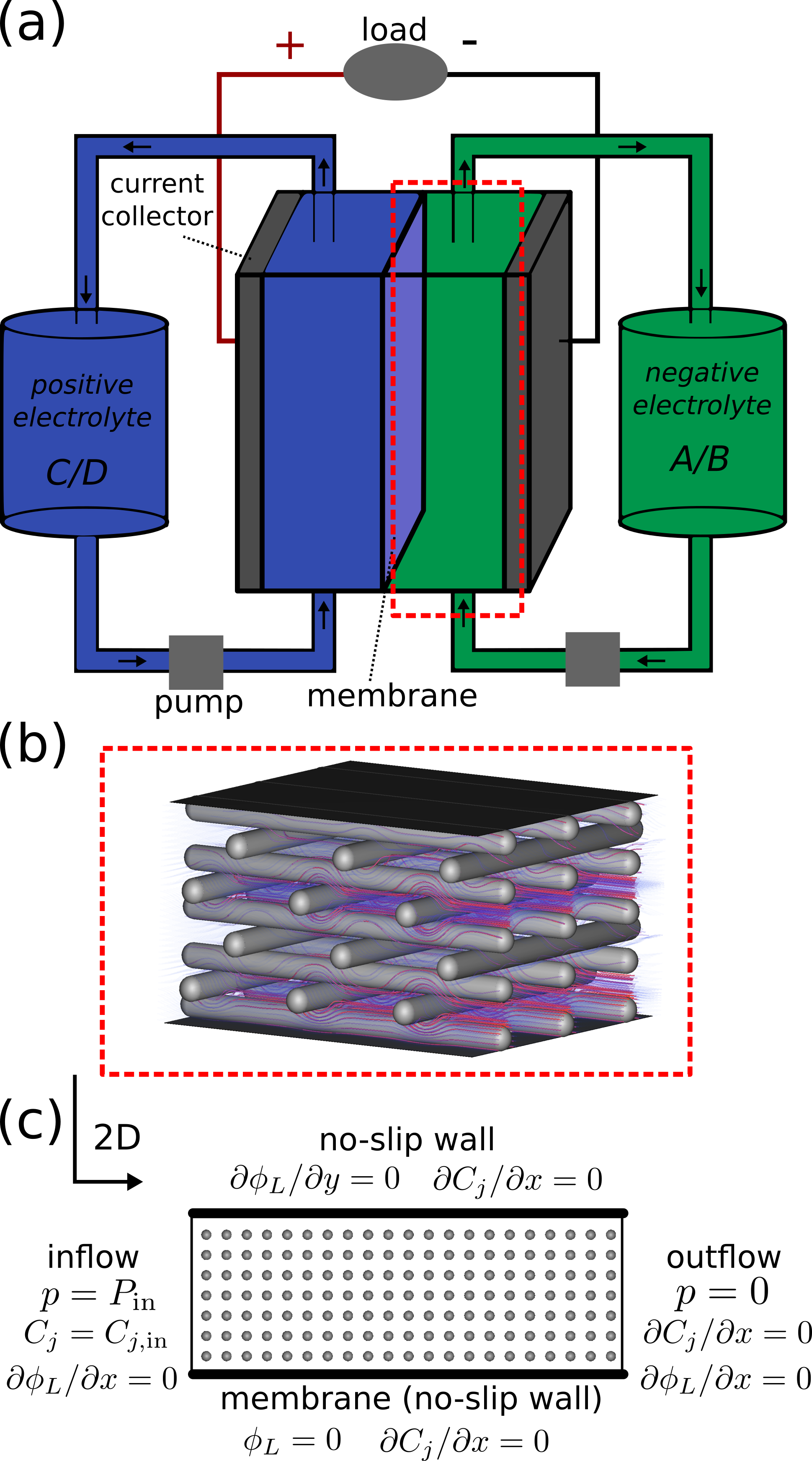}
\caption{Schematic of (a) full redox flow battery and (b) half-cell with an example of electrode microstructure (experimentally realizable with direct ink printing method). (c) Boundary conditions employed in this study. At the inlet, pressure $p$ and concentrations $C_j$ are fixed, while no flux boundary conditions are assumed for the electrolyte potential $\phi_L$. No-slip walls with no concentration flux are assumed at the bottom and top of the system. The bottom wall, where the membrane is located, is also the reference value (i.e.\@ the ground) for the electrolyte potential; whereas no flux condition for $\phi_L$ is imposed at the top wall. At the outlet, fixed pressure and no flux conditions for the other variables are imposed.}
\label{fig:schematic}
\end{figure}

A RFB is composed of two separate tanks, containing the liquids (electrolytes) in which the reactive species are dissolved, and a stack of electrochemical cells, comprising of two porous electrodes separated by a permeability-selective membrane or a porous separator. The electrolytes are pumped into the porous structures, where the redox reactions occur at the solid--liquid interfaces. Electronic current is generated in the solid, while ions flow in the liquid, with at least one charged species traversing the separator to pass charge internally between the two half-cells (Fig.~\ref{fig:schematic}). Depending on the applied load, current can be extracted or the inverse electrochemical processes can occur and the battery can be recharged. To improve RFB performance, several research avenues have been pursued, ranging from the development of new chemical compounds,~\cite{Er2015,Wu2021} to the optimization the microfluidic channel design of the cells,~\cite{Gerhardt2018} to several chemical~\cite{Yarlagadda2016}, thermal~\cite{Park2016} and physical~\cite{Mayrhuber2014} pre-treatments to enhance reactivity and conductivity of the solid surfaces. Recent studies focused on organic flow batteries, particularly quinone-based electrolytes, that often exhibit rapid and reversible redox reactions.~\cite{Huskinson2014,Lin2015,Er2015,Wu2021} When using a kinetically facile redox couple, the electrochemical performance (at the half-cell level) is ultimately determined by how the electrolyte flows within the porous electrode. However, a still unanswered question is ``what should the microstructure of the solid material look like to obtain a flow pattern that allows for a good battery performance?'' 

The need of filling this knowledge gap has become more evident in the last several years.~\cite{Zhou2017,Forner-Cuenca2019a} Recent experimental studies reported different performances when using various commercial electrodes made of carbon papers, cloths, or felts.~\cite{Forner-Cuenca2019, Tenny2020} These results suggested that a bimodal pore size distribution of the cloth microstructure was responsible of the superior performance of the cloth electrodes compared to the paper ones. Previously, experiments on laser-perforated carbon papers showed that higher current can be extracted when removing some of the material. It was hypothesized that the larger pores would act as highways for the reactive species, allowing for a more facile mass transport and for replenishing of the reactant.~\cite{Mayrhuber2014} Despite this empirical evidence, the intrinsic difficulties in experimentally characterizing the porous structures, especially \textit{in situ} and after they have been compressed to fit into the electrochemical cells, hinder the progress of correlating structural features with electrochemical performance. Furthermore, new synthetic routes are now appearing and opening up the possible design space of electrode microstructures, allowing to improve over the traditional commercial electrodes.~\cite{Wan2021,Beck2021} The theoretical identification of general design rules would substantially help these experimental efforts. Finally, large heterogeneities in the concentration profiles within the porous electrodes have been recently observed under operating conditions,~\cite{Wong2021} emphasizing the need to go beyond the traditional use of homogeneous models to investigate the role of microstructure on the electrochemical performance.

Distinct from homogeneous models~\cite{Gerhardt2018, Ryan2019, Chakrabarti2020} and pore-network simulations,~\cite{Sadeghi2019, GayonLombardo2019} pore-scale direct numerical simulations represent the natural tool to describe solid--liquid interfaces and obtain microscopic insights in the physical processes related to the electrode microstructure. Since such an approach is technically challenging and computationally expensive, studies of this kind are scarce. Pioneering pore-scale RFB simulations~\cite{Qiu2012, Qiu2012a} relied on the lattice Boltzmann (LB) method~\cite{Succi2001,Montemore2017} to obtain the velocity field within a porous material and subsequently solve the species and charge transport equations using a finite volume scheme. More recent works employed LB both for the flow and the transport simulations.~\cite{Chen2017, Ma2019, Ma2020, Zhang2018, Zhang2020} In Ref.~\citenum{Chen2017}, despite the electrolyte potential not being explicitly considered, multiphase LB was employed to gain insights on how oxygen bubbles evolution effectively reduces the available reactive surface of disordered electrodes used in vanadium RFBs. In Ref.~\citenum{Zhang2020}, three different microstructures were simulated and by introducing an experimentally calibrated fiber-level mass transport coefficient to better capture the effective reaction constant in a TEMPO-based RFB, good agreement with experimental data was obtained. All-copper~\cite{Ma2019} and iron-vanadium~\cite{Ma2020} RFBs have been also simulated.  Generally, however, the heavy computational burden of pore-scale simulations required the sacrifice of an accurate description of at least one of the solid--liquid interface (i.e.\@ using a crude binary discretization of either solid or liquid voxels); or the resolution of the simulation domain (e.g.\@ typically around \SI{3}{\um}--\SI{4}{\um} per voxel); or the system size. Overall, in these studies only few microstructures have been investigated and several structural parameters (e.g.\@ porosity, reactive surface area, fiber diameter, geometry) have been typically varied at the same time. As a result, precise design concepts associated to the electrode geometry are still missing. 

In this study, we perform direct numerical simulations of half-cell of a RFB (Fig.~\ref{fig:schematic}(b)) using a novel computational framework, efficiently parallelized to run both on standard workstations and on large computing clusters.~\cite{Zhang2019} Distinct from previous pore-scale simulations of RFBs, where each cubic voxel was discretized either as solid or liquid, our method employs a cut-cell approach, where voxels can also be cut by the boundary between solid and liquid, resulting in a more accurate description of the interfaces. Furthermore, fluid incompressibility is naturally enforced in our simulations compared to previous LB-based studies. By sacrificing one spatial dimension and considering idealized two-dimensional (2D) geometries, we are able to directly simulate systems that are longer along the flow direction than considered before and show non-trivial effects caused by specific features of the electrode microstructure. In particular, inspired by the possibility of precisely controlling microstructure through additive manufacturing such as direct ink printing, we focus on how a regular lattice can be modified to improve electrode performance. Contrary to intuition, we observe that by removing reactive material, i.e.\@ reducing the amount of liquid--solid interface, it is possible to extract higher current, since mixing of the reactive species is favored. By carefully engineering the location of vacancies, we show that it is possible to outperform disordered structures. This suggests a way toward designs that are superior to current commercial electrode microstructures.

\section{\label{sec:theo}Theoretical background}
We perform direct simulations of fluid flow and advection--diffusion--electromigration--reaction within porous structures. Our approach consists in (i) obtaining the flow profile of an incompressible Newtonian fluid by solving the Navier--Stokes equations, and (ii) solving the coupled transport and electrochemical reaction equations to obtain concentration profiles and electrolyte potential. The governing equation for the first part of the simulations is
\begin{equation}
\frac{\partial \mathbf{u}}{\partial t} = - \mathbf{u} \cdot \nabla \mathbf{u} + \frac{1}{\rho} \left( -\nabla p + \mu \nabla^2 \mathbf{u} \right) \,\, ,
\end{equation}
where $\rho$ is the fluid density, $\mu$ is the dynamic viscosity, $p$ is the pressure, and $\mathbf{u}$ is the velocity field subject to the incompressibility condition
\begin{equation}
\nabla \cdot \mathbf{u} = 0 \,\, .
\end{equation}
No-slip boundary conditions are assumed at the solid--liquid interfaces. The system is evolved in time until steady-state behaviour is reached. The flow velocity is considered at steady state when the $\ell_2$ norm of the relative velocity difference between two time steps is less than a certain threshold ($\epsilon_0=10^{-5}$):
\begin{equation}
\frac{\|\mathbf{u}_{n} - \mathbf{u}_{n-1} \|}{\| \mathbf{u}_{n-1}\|} < \epsilon_0 \,.
\end{equation}
The obtained velocity profile is used as input for the second part of the simulations.  We assume that variations in the species concentration do not influence the flow, e.g.\@ the viscosity remains constant, and therefore there is a one-way coupling from the fluid problem to the mass--charge transport problem. Under the common assumption of dilute solution, the transport equation for the species $j$ reads as
\begin{equation}
\label{eq:ader}
\frac{\partial C_j}{\partial t} = D_j \nabla^2 C_j - \mathbf{u} \cdot \nabla C_j + \frac{z_j D_j F}{RT} \nabla \cdot \left( C_j \nabla \phi_L \right) + S_j
\end{equation}
where $C_j$ is the local concentration, $D_j$ is the diffusivity, $z_j$ is the charge, $S_j$ is the reaction (source or sink) term, $\phi_L$ is the electrolyte potential, $F$ is Faraday's constant, $R$ is the ideal gas constant, and $T$ is the temperature (assumed at the constant value of $\SI{298}{K}$). Note that the electromigration term, which is third on the right-hand side of Eq.~\eqref{eq:ader} is not neglected here, different from previous works.~\cite{Qiu2012,Qiu2012a} In this study, we consider the discharging of a quinone-based RFB.~\cite{Huskinson2014} The redox reaction between species A (to indicate the hydroquinone H$_2$AQDS) and B (AQDS) involving $\nelec=2$ electrons and $\nprot=2$ protons can be written as
\begin{equation}
  \text{A} \rightarrow \text{B} + \nelec \text{e}^- + \nprot \text{H}^+  \,\, .
\end{equation}
Dispersed in the fluid, there is also a large amount of $\text{H}_2\text{SO}_4$ ($C_{\text{H}_2\text{SO}_4}^{(\text{diss})}=\SI{1}{\mole/\liter}$), acting as supporting electrolyte, which we assume to be dissociated only in $\text{H}^+$ and $\text{HSO}_4^-$. Because of the electroneutrality condition, only the transport equations for the species $j=\{\text{A}, \, \text{B}, \, \text{H}^+\}$ are explicitly solved, whereas the concentration of $\text{HSO}_4^-$ is obtained via $\sum_l z_l C_l =0$, where $l=\{\text{A}, \, \text{B}, \, \text{H}^+, \, \text{HSO}_4^-\}$. The local production/depletion of chemical species is coupled with the potential difference between the liquid electrolyte and the solid electrode via the Butler--Volmer equation. The reaction term is therefore defined as
\begin{equation}
\label{eq:source}
S = a k \, {C_\text{A}}^{\alpha_\text{A}} \, {C_\text{B}}^{\alpha_\text{B}}\,
\left[\; \exp \left(\frac{\nelec F \alpha_\text{A}}{RT} \eta \right)- \exp \left(-\frac{\nelec F \alpha_\text{B}}{RT} \eta \right) \; \right] \,\, ,
\end{equation}
where $a$ is the specific area of the solid--liquid interface, $k$ is the reaction constant, and $\eta$ is the local overpotential. $\alpha_\text{A}$ and $\alpha_\text{B}$ are the anodic and cathodic charge transfer coefficients; we assume a perfectly reversible reaction for which $\alpha_\text{A}=\alpha_\text{B}=0.5$. The appropriate stochiometric prefactors---to obtain $S_j$ of Eq.~\eqref{eq:ader} from $S$ defined above---are $-1,+1,\nprot$ for $\text{A}$, $\text{B}$, and $\text{H}^+$, respectively. The overpotential is defined as
\begin{equation}
\label{eq:overpot}
  \eta=  \phi_S-\phi_L-E_{eq} \, ,
\end{equation}
with $\phi_S$ being the potential in the solid, and the Nernst equilibrium potential given by~\cite{Knehr2011}
\begin{equation}
\label{eq:eeq}
E_{eq}=E_0 + \frac{RT}{\nelec F}\log\frac{C_\text{B} (C_{\text{H}^+})^{\nprot}}{C_\text{A}} \, .
\end{equation}
The electrolyte potential is obtained by solving
\begin{equation}
\label{eq:potl}
- \kappa_{L} \, \nabla^2 \phi_L - F \, \sum_l \, z_l \, D_l \, \nabla^2 C_l \, = - S_{\phi} \, ,
\end{equation}
where $l$ runs over all species, the source term is given by $S_{\phi}=\nelec F S$, and the electrolyte conductivity depends on the local concentrations as
\begin{equation}
\label{eq:kappa}
\kappa_{L} = \frac{F^2}{RT} \sum_l z_l^2 D_l C_l \, .
\end{equation}
In this study, we perform potentiostatic simulations and, by assuming high conductivity in the solid, we consider a constant electric potential throughout the solid $\phi_S = V_{\text{app}}$. 
Our goal is to focus on the performance losses associated to the electrode microstructure and we therefore simulate only an half-cell, neglecting the transport within the membrane and the associated potential drop. 
By using the bottom wall, where the membrane is located, as potential ground (for which $\phi_L=0$), we can manipulate Eq.~\eqref{eq:eeq} to incorporate the constant terms in this reference value and set
\begin{equation}
E_0^{'}= E_0+\frac{\nprot RT}{\nelec F}\log  C_{\text{H}^+}^{(0)}=0 \, ,
\end{equation}
where $C_{\text{H}^+}^{(0)}$ is the initial proton concentration. We therefore use 
\begin{equation}
\label{eq:eeqprime}
E_{eq}^{'}=\frac{RT}{\nelec F}\log\frac{C_\text{B}}{C_\text{A}} +\frac{\nprot RT}{\nelec F}\log  \left( 1+ \frac{C_{\text{H}^+}-C_{\text{H}^+}^{(0)}}{C_{\text{H}^+}^{(0)}} \right)
\end{equation}
to calculate the local Nernst equilibrium in our simulations. In this way, $V_{\text{app}}$ can be viewed as a half-cell overvoltage, in which membrane losses are not included. Finally, we note that in this study we do not employ any additional equation involving a mass transfer coefficient from the bulk to the fiber surface. Since we use a fine mesh resolution, the concentrations appearing in Eqs.~\eqref{eq:source} \&~\eqref{eq:eeqprime} are the ones at the cell center, in line with previous studies.~\cite{Qiu2012,Qiu2012a} Introducing a fiber-level mass transfer coefficient would decrease the species concentration participating in the reaction and the corresponding local reaction rate. It could be useful to capture e.g.\@ the dependence on fiber roughness and local velocity, but this is beyond the scope of this study. 

The boundary conditions are shown in Fig.~\ref{fig:schematic}(c), and consist of top and bottom impermeable no-slip walls, and an inlet on the left, where uniform concentrations are specified, and an outlet on the right. The flow is driven by a pressure difference $\Delta P$ between inlet and outlet. Characteristic operational regimes of RFBs have small but finite Reynolds number and very large Schmidt and P\'eclet numbers. Here we fix the physical parameters of the electrolyte based on the 9,10-anthraquinone-2,7-disulphonic acid (AQDS) RFB assuming a reversible reaction and a symmetric redox couple (see Table~\ref{tab:params}). 
The composition of the electrolyte at the inlet is fixed at an initial state of charge (SOC), that is the ratio between product and reactant, $\text{SOC}^{(0)}=C_\text{B}^{(0)}/(C_\text{A}^{(0)}+C_\text{B}^{(0)})=0.5$ and $C_\text{A}^{(0)}=C_\text{B}^{(0)}=\SI{0.25}{\mole/\liter}$. Correspondingly, the proton concentration at the inlet is $C_{\text{H}^+}^{(0)}=C_{\text{H}^+}^{(\text{diss})}+\nprot C_\text{B}^{(0)}=\SI{1.5}{\mole/\liter}$.
Different operating conditions are explored by varying the flow rate (controlled by $\Delta P$) and the applied (half-cell) overvoltage $V_{\text{app}}$. We define the current density extracted from the simulated (2D) half-cell $j$ as the total current divided by the membrane area (see below for a more accurate definition). When comparing different structures we will consider simulations performed at the same $\Delta P$. This automatically corresponds the same pumping losses, the cost to push the electrolyte through the porous electrode. Therefore, the microstructure exhibiting higher current is directly the one with superior battery performance. By performing 2D simulations we aim to illustrate the influence of several structural features on the battery performance. A quantitative comparison between realistic three-dimensional (3D) microstructures is left for future study.

\begin{table}[h!]
  \caption{\label{tab:params}Simulation parameters. Electrolyte properties based on Ref.~\citenum{Huskinson2014}.}
\begin{ruledtabular}
\begin{tabular}{llll}
Quantity & Symbol & Value & Units\\
\hline
Fluid density & $\rho$ & $1000^\text{\strut}$ & \si{\kg/\m^3}\\
Fluid viscosity & $\mu$ & $10^{-3}$ & \si{\newton \second / \m^2}\\
Diffusivity reactant & $D_{\text{H}_2\text{AQDS}}$ & $4\times 10^{-10}$ & \si{\metre^2 / \second}\\
Diffusivity product  & $D_{\text{AQDS}}$ & $4 \times 10^{-10}$ & \si{\metre^2  / \second}\\
Diffusivity ions & $D_{\text{H}^+}$ & $9.3 \times 10^{-9}$ & \si{\metre^2 / \second}\\
Diffusivity counterions & $D_{\text{HSO}_4^-}$ & $1.38 \times 10^{-9}$ & \si{\metre^2 / \second}\\
Charge reactant & $z_{\text{H}_2\text{AQDS}}$ & $-2$ & --\\
Charge product  & $z_{\text{AQDS}}$ & $-2$ & --\\
Charge ions & $z_{\text{H}^+}$ & $+1$ & --\\
Charge counterions & $z_{\text{HSO}_4^-}$ & $-1$ & --\\
Reaction constant & $k_0$ & $7.2 \times 10^{-5}$ & \si{\metre / \second}\\
Transfer coefficients & $\alpha_\text{A}$, $\alpha_\text{B}$ & 0.5 & --\\
Initial concentration reactant & $C_{\text{H}_2\text{AQDS}}^{(0)}$ & 0.25 & \si{\mole / \liter}\\
Initial concentration product & $C_{\text{AQDS}}^{(0)}$ & 0.25 & \si{\mole / \liter}\\
Initial concentration ions & $C_{\text{H}^+}^{(0)}$ & 1.5 & \si{\mole / \liter}\\
\end{tabular}
\end{ruledtabular}
\end{table}

\section{\label{sec:num}Computational method}
Our simulation code is built within the AMReX framework,~\cite{Zhang2019} and it is based on a recent open-source AMReX-based fluid solver (``incflo'').~\cite{Sverdrup2019} It employs a finite-volume approach with a projection method to deal with the incompressible Navier--Stokes equations,~\cite{chorin68,Almgren1998} and an embedded boundary (EB) description for the solid--liquid interfaces.~\cite{Johansen1998, Graves2013, Trebotich2015} We extended this code to deal with the second part of our simulations, i.e.\@ advection--diffusion--electromigration--reaction of chemical species coupled with the electrolyte potential equations. Within the EB approach, the source term is non-zero only in cut-cells, where quantities associated to both the solid and the liquid are defined. The specific reactive area appearing in Eq.~\eqref{eq:source} depends on the local geometry information as $a_i=A_i^\text{EB} / V_i= \beta^\text{EB}_i / \Delta x$, where $A^\text{EB}_i$ is the area of the solid surface in the cut-cell $i$, $V_i$ is the cell volume,  $\beta^\text{EB}_i$ is the solid area fraction, and $\Delta x$ is the regular cell dimension (or mesh resolution). The definition of the current density is related to the total current $I_\text{tot}$ and reads
\begin{multline}
j= \frac{I_\text{tot}}{L_x L_z} = \frac{1}{L_x L_z} \sum_i j_i \, A^\text{EB}_i = \frac{1}{L_x L_z} \sum_i \, j_i \beta^\text{EB}_{i} \, \Delta x \, L_z =\\
\frac{1}{L_x} \sum_i \nelec F k C_{A,i}^{\alpha_A} C_{B,i}^{\alpha_B} \left[ e^{\left( \frac{\nelec F \alpha_A}{RT} \eta_i \right)} - e^{\left( -\frac{\nelec F \alpha_B}{RT} \eta_i  \right)} \right] \, \beta_i^\text{EB} \Delta x \, ,
\end{multline}
where the sum extends to all the cut-cells $i$, $j_i$ is the local current density (with the quantities defined as in the previous section), $L_x$ is the system length along the flow direction. $L_z$ refers to the third dimension that is not simulated, but it cancels out in this definition of $j$. Note that $j$ has units of \si{\ampere/\m^2}, despite the simulations being performed in 2D.

\begin{algorithm}
\DontPrintSemicolon
\label{method}
\Do{$\frac{\|C_j^{(n+1)}-C_j^{(n)}\|}{\Delta t \| C_j^{(n)}\|}< \epsilon_3$
and $\frac{\|\phi_L^{(n+1)}-\phi_L^{(n)}\|}{\Delta t \| \phi_L^{(n)}\|}< \epsilon_4$}{
    Time step $n \rightarrow n+1$ \;
    Store $C_j^{(n)}, \phi_L^{(n)}$ \;    
    $C_j^{(n,*)}=C_j^{(n)}+\Delta t \mathcal{A}_j^{(n)}$ \;
    Set $k=0$, $C_j^{(k=0)}=C_j^{(n,*)}$, $\phi_L^{(k=0)}=\phi_L^{(n)}$ \;
    \Do{$\frac{\|C_j^{(k+1)}-C_j^{(k)}\|}{\| C_j^{(k)}\|}< \epsilon_1$
and $\frac{\|\phi_L^{(k+1)}-\phi_L^{(k)}\|}{\| \phi_L^{(k)}\|}< \epsilon_2$}{
        Fixed point iterator $k \rightarrow k+1$ \;
        Calculate $S^{(k)}$, $\mathcal{E}^{(k)}$, ${\kappa_L}^{(k)}$, $\mathcal{F}^{(k)}$ \;
        Solve $(1-\Delta t D_j \nabla^2)C_j^{(k+1)}=C_j^{(n,*)}+\Delta t (\mathcal{E}^{(k)} + S_j^{(k)})$\;
        Solve $-\nabla \cdot (\kappa_L^{(k)} \nabla) \phi_L^{(k+1)}=\mathcal{F}^{(k)}- S_{\phi}^{(k)}$ \;
        $C_j^{(k+1)}=(1-\omega)C_j^{(k)} + \omega C_j^{(k+1)} $ \;
        $\phi_L^{(k+1)}=(1-\omega)\phi_L^{(k)} + \omega \phi_L^{(k+1)} $ \;
        }
    }
\caption{Time-stepping algorithm for the coupled mass and charge transport problem. $\mathcal{A}$ and $\mathcal{E}$ are the advection and electromigration terms in Eq.~\eqref{eq:ader}. $\mathcal{F}$ is the ionic flux in Eq.~\eqref{eq:potl}.}
\end{algorithm}

Time-stepping is performed to reach the steady-state solution. To advance from step $n$ to step $n+1$, we use an operator splitting approach, where first the advection term (denoted as $\mathcal{A}$) is treated explicitly in the same fashion as in Ref.~\citenum{Sverdrup2019}, where a flux-based redistribution scheme is used for the cut-cells. The other terms (diffusion, electromigration, and reaction) are considered semi-implicitly using a fixed-point iteration method. The procedure is schematically shown in Alg.~\ref{method}. Given the typical operating conditions of an RFB and the mesh resolution employed, the advection term globally dominates the species transport. Therefore, the time increment $\Delta t$ is based on the Courant--Friedrichs--Lewy (CFL) condition imposed by the advection term (with a typical CFL coefficient of 0.75). To ensure the stability of the iterative fixed-point method in case of strong non-linearities (e.g.\@ for large applied potential), a relaxation step is employed, where the quantities are updated with a relaxation factor $\omega$ that is set to values smaller than 1. The systems of linear equations resulting from the various steps can be expressed as $(\alpha A - \beta \nabla \cdot B \nabla) \psi=f$, where $\alpha$ \& $\beta$ are scalar constants, $A$ \& $B$ are scalar fields, $\psi$ is the unknown and $f$ is the right-hand side. These were efficiently solved by exploiting both the solvers built into the AMReX package, and the HYPRE library.~\cite{hypre} In particular, we found that to solve the diagonally-dominated equations ($A,\alpha \neq 0$), e.g.\@ arising from the implicit diffusion steps, the AMReX native solvers (geometric multigrid and BiCGStab) are efficient methods. However, to solve the equations with $\alpha=0$ needed for the MAC projection step, for the pressure projection step, and for obtaining the electrolyte potential, we employed the GMRES algorithm preconditioned with the algebraic multigrid BoomerAMG from the HYPRE library. Similar observations were made for other EB-based reactive transport simulations.~\cite{Trebotich2014} Results obtained with our simulation framework agree with literature benchmarks for reactive fluids (in the context of static mineral dissolution).~\cite{Molins2020} 

\begin{figure*}
\includegraphics[width=0.80\textwidth]{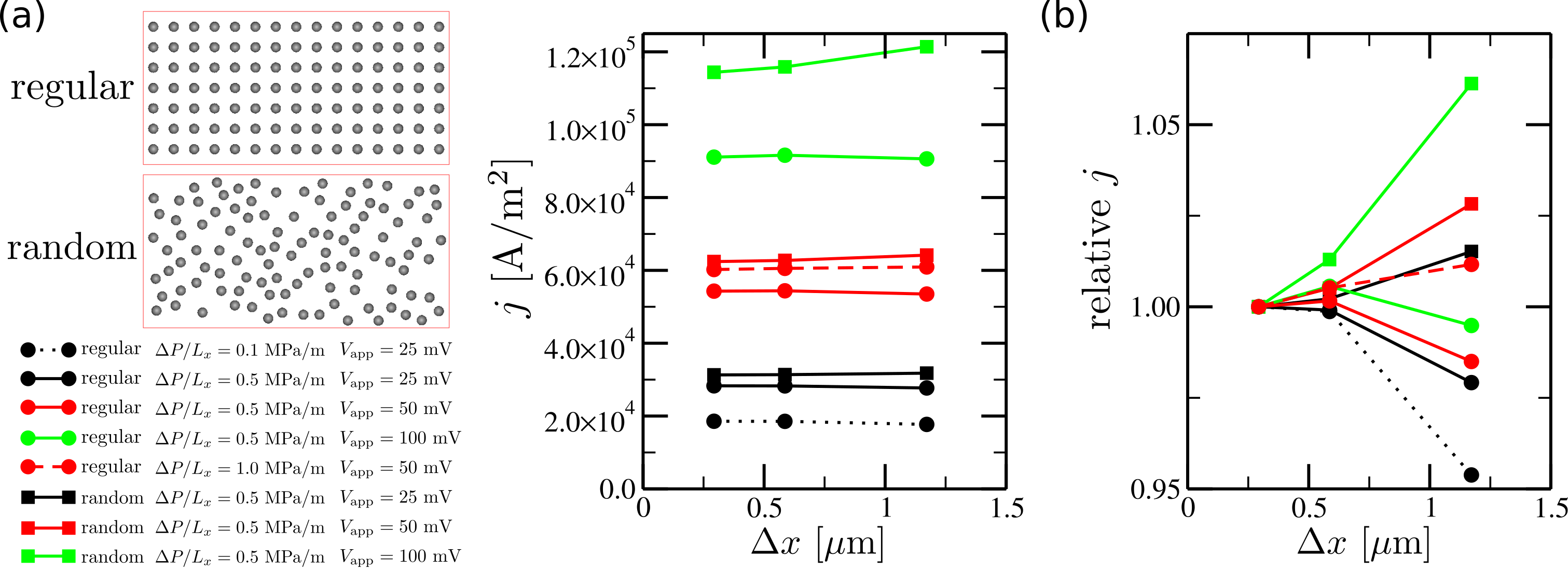}
\caption{(a) Mesh convergence study for regular (ordered) and random (disordered) configurations ($L_x=\SI{300}{\um}$) simulated at various grid resolution $\Delta x$ under different operating conditions. (b) Relative current density with respect to the finest grid resolution. We observe that a resolution of $\Delta x \simeq \SI{0.58}{\um}$ is adequate for this study.}
\label{fig:meshres}
\end{figure*}

To identify a suitable grid resolution for our investigation, we performed several simulations at different mesh resolution $\Delta x$ for a range of typical RFB operating conditions (imposed pressure and potential) that will be considered in this study. The results obtained for a regular (square lattice) and a disordered configuration of $N=$ cylinders are reported in Fig.~\ref{fig:meshres}. In both cases, the system dimensions are $L_x=\SI{300}{\um}$ and $L_y=\SI{150}{\um}$. From this analysis we observed that results for $\Delta x \simeq \SI{0.58}{\um}$ are within less than 1.5\% (in the worst case) from the more expensive finer resolution. We therefore consider such resolution adequate for our purposes and it will be employed in the simulations of larger systems. Our study consists of a systematic comparison between different configurations of identical objects (cylinders). Therefore, to ensure that every cylinder is described in the same way within the EB algorithm, we imposed cylinder positions to be at centers of the grid cells.
In this way, at fixed mesh resolution, porosity and reactive surface area of configurations with same number of cylinders arranged in different fashion, are still kept equal. 

\section{\label{sec:res}Results and Discussion}

\begin{figure*}
\includegraphics[width=0.99\textwidth]{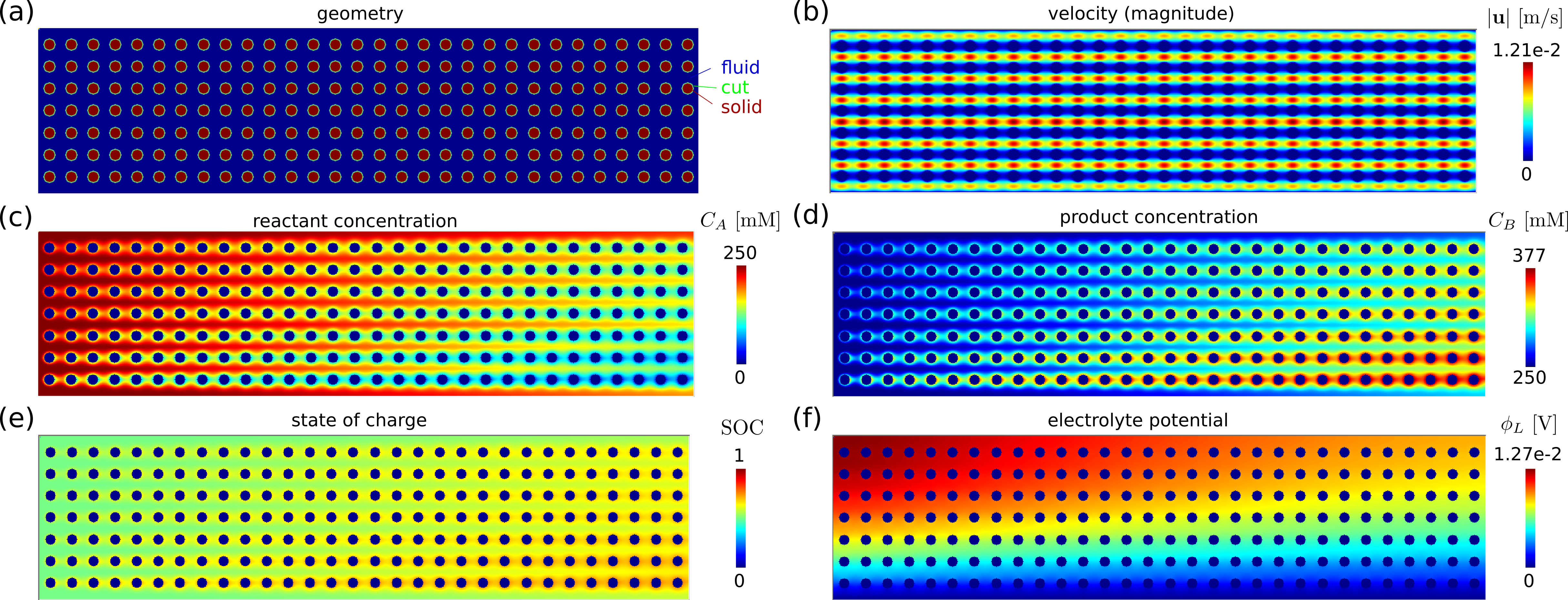}
\caption{Typical simulation output for a regular lattice configuration ($L_x=\SI{600}{\um}, \Delta P/L_x=\SI{0.5}{\MPa/\m}, V_{\text{app}}=\SI{25}{mV}$). The lattice geometry is discretized according to the embedded boundary method (a) and the flow velocity at steady-state is shown in (b). The concentration profiles (c-d) and the associated state of charge (e) show how the material upstream creates trails of product investing the subsequent cylinders (shadowing effect). In (f) the potential gradient that develops from the membrane (bottom) to the top of the half-cell is shown.}
\label{fig:example}
\end{figure*}

\subsection{Transport in a RFB lattice electrode}

To briefly discuss the main transport phenomena occurring in a RFB, in Fig.~\ref{fig:example} we show the typical output of our simulation framework for the case of a regular lattice, that is also the starting point of our investigation. The configuration consists of 210 cylinders with a diameter of \SI{10}{\um} arranged in a square lattice of 7 rows with center-to-center spacing of \SI{20}{\um}. The system is confined between two walls, with the cylinders in the first and last row at a distance of \SI{15}{\um} from the walls, resulting in a porosity of $\epsilon \simeq 0.817$. Panel (a) shows the discretization with the EB method. The system dimensions are $L_x=\SI{600}{\um}$ and $L_y=\SI{150}{\um}$, the latter being in the range of thin and compressed commercial electrodes. During our design study, $L_y$ will be kept fixed and we will vary the arrangement of the cylinders. The operating conditions are defined by $\Delta P$, which controls the flow rate, and by $V_{\text{app}}$, a proxy for the half-cell (over)potential (excluding membrane losses). In Fig.~\ref{fig:example}, we report the main quantities at steady-state for $\Delta P/L_x=\SI{0.5}{\MPa/\m}$ and $V_{\text{app}}=\SI{25}{\mV}$. Panel (b) shows the flow speed, which is highest in between the objects in the middle of the channel and zero at the walls (where no-slip boundary conditions are applied). Panels (c) and (d) show the reactant and product concentration, respectively. The reactant concentration is decreasing along the flow direction $x$, whereas the product concentration is increasing. The proton concentration (not shown) follows a qualitatively similar decrease along $x$, however quantitatively it is less affected due to the much higher diffusivity of $\text{H}^+$. The state of charge is shown in panel (e). For all simulations, the initial SOC is 0.5. The electrolyte potential is shown in panel (f), from which a gradient from bottom to top is clearly observed. The potential ground is defined such that $\phi_L=0$ at the bottom, where the membrane is located. Since the potential difference $\Delta \phi=\phi_S-\phi_L$ is a key term driving the reaction (Eqs.~\eqref{eq:source} \& \eqref{eq:overpot}), the reaction rate is higher close to the membrane due to the presence of the gradient. Therefore, the increase in SOC is larger in the lower rows of cylinders. From the concentration profiles, it is evident that due to the reaction, the objects upstream generate a trail of fluid with different SOC that invest the material downstream. This shadowing phenomenon results in a higher SOC on the surface of the downstream cylinder. Correspondingly, the concentration overpotential, that is the part of $\eta$ related to the concentrations (cfr. Eqs.~\eqref{eq:overpot} and~\eqref{eq:eeq}), decreases. Shadowing is therefore the cause of a smaller reaction rate. Due to the coupling between mass and charge transport, $\phi_L$ is also partially affected and in fact its magnitude slowly decreases along the x direction. The overall result is that less current can be extracted from the material downstream due to the shadowing effect. This can be also interpreted as an example of fuel starvation. The lattice performance is summarized by polarization curves, shown in Fig.~\ref{fig:regvac}(a) with solid lines for this configuration, where the extracted current density $j$ is plotted as a function of $V_{\text{app}}$ for different $\Delta P$. After an initial linear increase, the curves start to level off towards a limiting current value, indicating that mass transport limitations are becoming more severe. By increasing the flow rate, this can be postponed to larger voltages and $j$ becomes higher.

\subsection{Less can be more}
\begin{figure}
\includegraphics[width=0.49\textwidth]{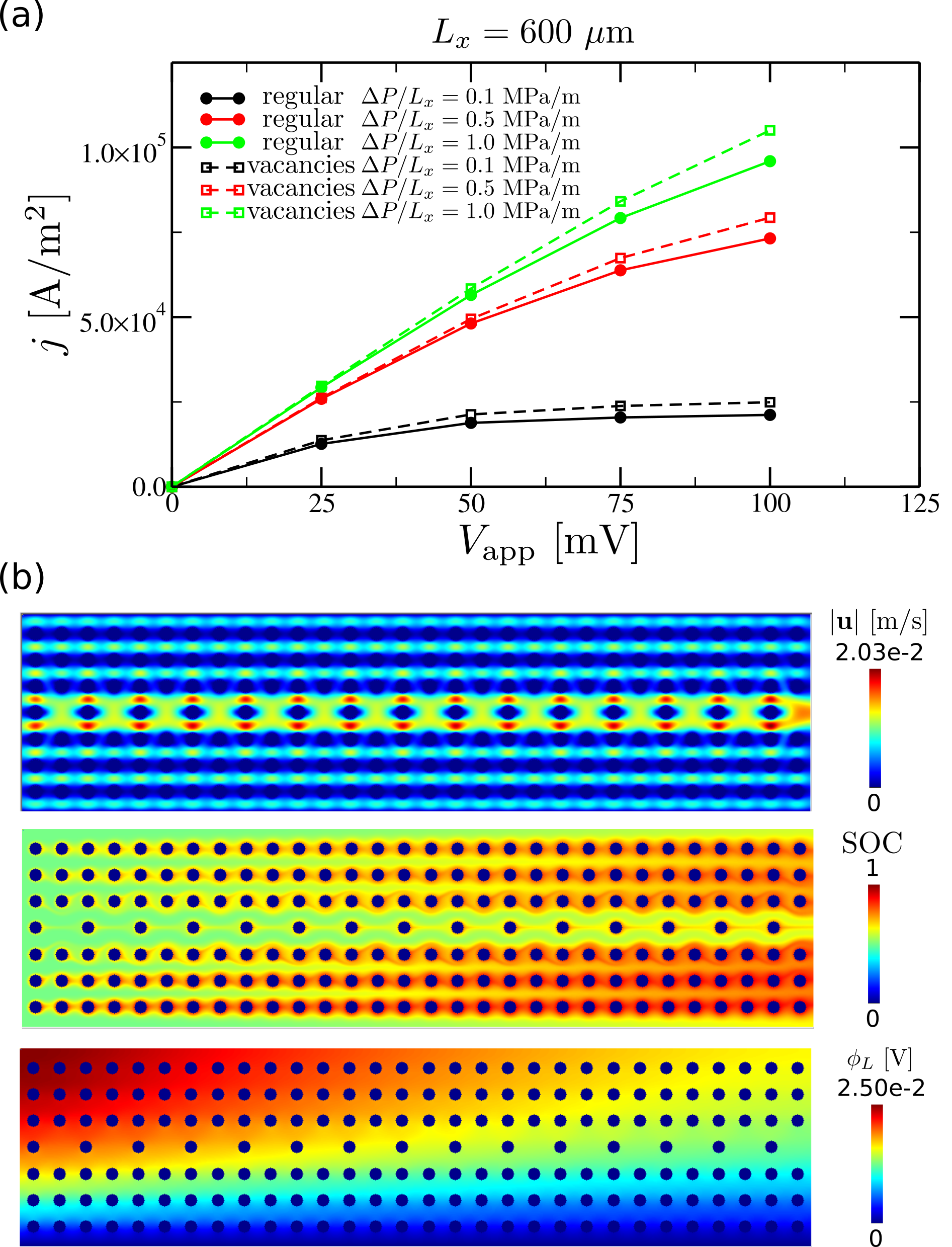}
\caption{(a) Current density as a function of applied potential for regular configurations and ones with vacancies in the middle row at different flow rates. (b) Associated snapshots of the simulated configuration with vacancies ($\Delta P/L_x=\SI{0.5}{\MPa/\m}$, $V_{\text{app}}=\SI{50}{\mV}$) showing velocity, state of charge and electrolyte potential. The configuration with vacancies outperforms the regular intact material, because of better mass transport properties despite featuring less reactive surface.}
\label{fig:regvac}
\end{figure}

The first variation with respect to the base case of a fully ordered lattice that we consider is the same lattice with vacancies in the middle row. In particular, one out of every two cylinders are removed from row number 4 as counted from the bottom, where the membrane is located. Intuitively, it might be expected that by reducing the total reactive surface area, the overall performance will decrease. Surprisingly, however, a higher current can be obtained by the configuration with vacancies, as shown in Fig.~\ref{fig:regvac}(a). The gain in performance is larger at higher applied voltages, where mass transport limitations typically kick in, but also at larger $\Delta P$, i.e.\@ by increasing the flow rate. As we can observe from the snapshots of Fig.~\ref{fig:regvac}(b), the removal of the objects has several consequences. First, the permeability of the system increases, implying that at the same $\Delta P$ (so at the same cost of pumping operation), the configuration with vacancies exhibits an overall faster flow. This is particularly true in the middle row where the flow speed in pores between objects can be almost twice as fast compared to the regular lattice (cfr.\@ top panel of Fig.~\ref{fig:regvac}(b) with Fig.~\ref{fig:example}(b)). Since generally higher current can be extracted at faster flow rate, the configuration with vacancies will outperform more and more the regular ones upon increasing $\Delta P$. Second, in the middle row the shadowing effect is attenuated, i.e.\@ the product trail (as evident from the SOC simulation snapshot) becomes thinner since there is a larger distance over which the reactive species can diffuse away from the main velocity streamline. As a consequence, the concentration overpotential in the downstream material is lowered. Third, the velocity field in the rows next to the middle (number 3 and 5) is perturbed by the presence of larger pores in row 4, therefore the downstream material in these rows is not fully invested by the trail generated by the object upstream. Also in this case, the concentration overpotential is therefore decreased. Both effects become more and more relevant upon increasing $V_{\text{app}}$. Finally, the potential gradient from the membrane to the top wall is still present and quantitatively is only slightly affected compared to the regular case (results not shown).

\begin{figure}
\includegraphics[width=0.49\textwidth]{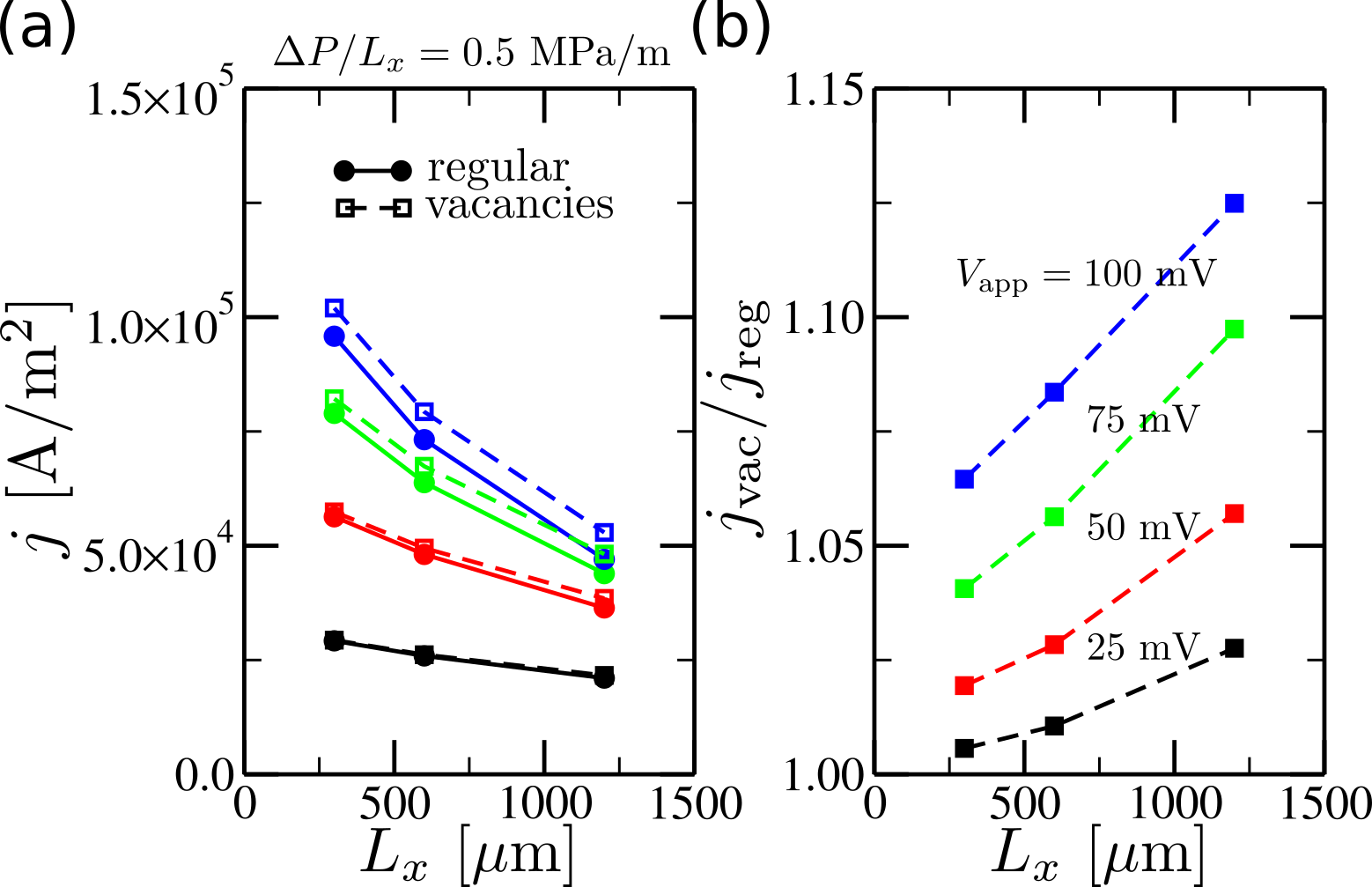}
\caption{Role of system size. (a) Current density as a function of system length for configurations with and without vacancies for four applied potentials $V_\text{app} = \SI{25}{mV}, \SI{50}{mV}, \SI{75}{mV}, \SI{100}{mV}$---see labels in panel (b) for a guide to the colors. (b) Increase in performance of configurations with vacancies with respect to regular ones as a function of $L_x$. Vacancies are more beneficial in longer systems where mass transport limitations are more severe for the same imposed pressure difference per unit length $\Delta P/L_x$.}
\label{fig:syssize}
\end{figure}

To investigate whether the increased performance could be due to finite-size or boundary effects, we simulate the same arrangements for systems with different $L_x$, while keeping $\Delta P/L_x$ constant, up to $L_x=\SI{1.2}{\mm}$, which is a directly comparable size for some experimental electrodes (e.g.\@ in interdigitated flow fields~\cite{Kumar2016,Messaggi2018,Gerhardt2018,Tsushima2020}). As shown in Fig.~\ref{fig:syssize}(a) the current density $j$ decreases upon increasing $L_x$, as expected since overall mass transport limitations become more severe as the system becomes longer (fuel starvation towards the end of the system). Interestingly, upon increasing $L_x$ the performance gain for the configuration with vacancies becomes larger (Fig.~\ref{fig:syssize}(b)). For both configurations, it would be possible to overcome fuel starvation in long systems and extract larger $j$ by increasing $\Delta P$. In such a regime, the configuration with vacancies would outperform the regular one, since it exhibits a more favorable trend with increasing $\Delta P$, as discussed earlier. We can therefore conclude that the increase in performance is not due to finite-size effects, and it would become more significant in longer systems. These are however more computationally expensive to be systematically simulated, both because of a larger mesh and a larger number of timesteps to reach the steady-state in such advection-dominated systems. In the following, we will then explore more configurations with $L_x=\SI{600}{\um}$.

\subsection{Engineering vacancies}

\begin{figure}
\includegraphics[width=0.49\textwidth]{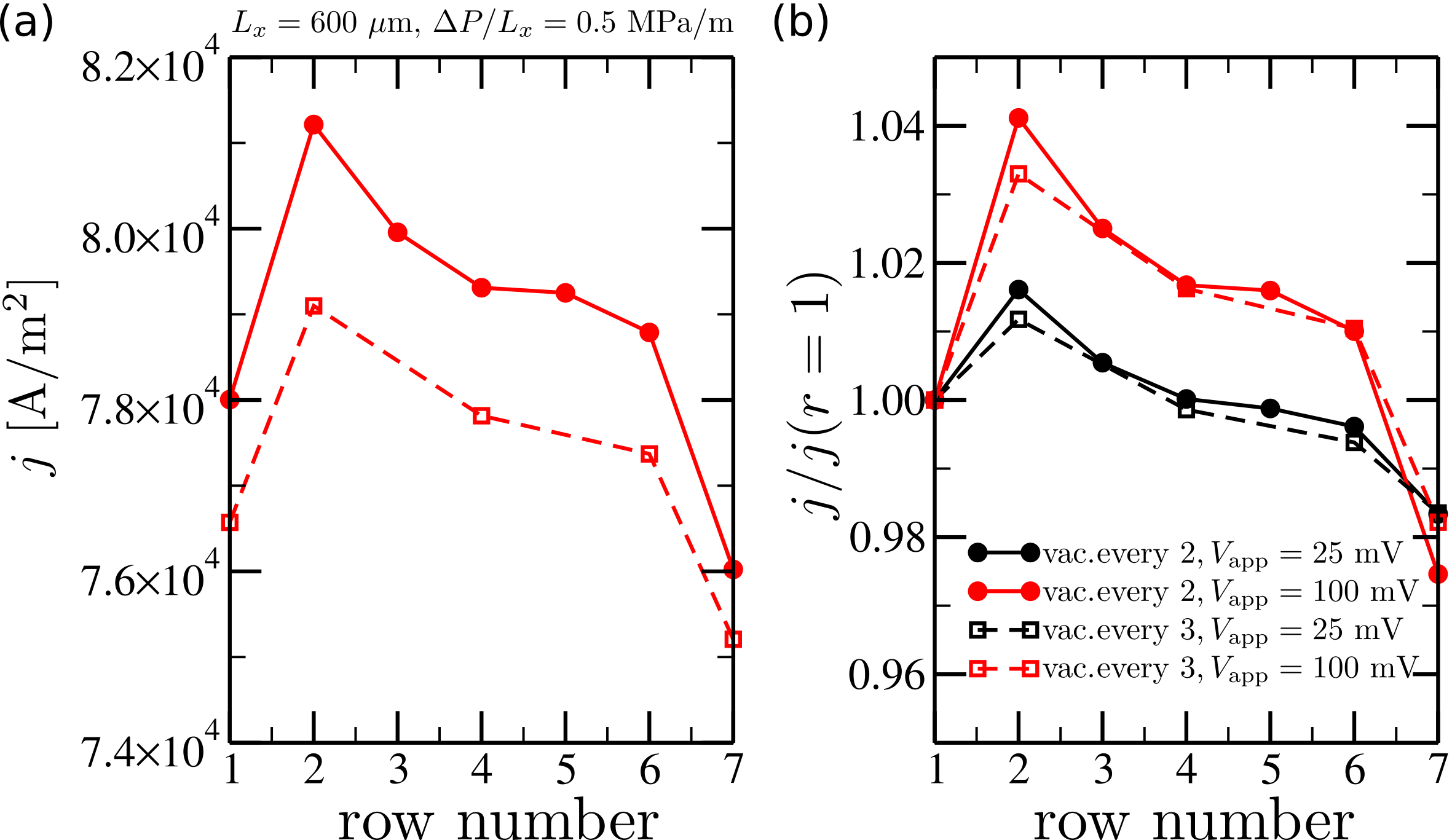}
\caption{(a) Current density as a function of the row featuring the vacancies, for configuration with vacancies every two cylinders (solid symbols) or every three (empty). Rows are numbered from the bottom (1=membrane) to the top wall. $V_{\text{app}}=\SI{100}{mV}$. (b) Relative current density normalized by the values of the configurations with vacancies in the first row. As a consequence of the interplay between shadowing effects and gradient in the electrolyte potential, a maximum is observed when vacancies are in the second row.}
\label{fig:vacrows}
\end{figure}

Due to the presence of hard walls at the top and the bottom of the system, placing the vacancies in a row different from the middle one would affect the velocity field in a different way. Additionally, since there is a potential gradient from the bottom to the top of the system, it can create an even stronger influence on how the locations of the vacancies quantitatively affect the concentration profiles. For these reasons, we simulated configurations still with vacancies in only one row, but now systematically varying such row. In Fig.~\ref{fig:vacrows}(a), we report the current density as a function of the row number, and clearly observe a non-monotonic trend, with a maximum for vacancies located in row 2. Placing vacancies in row 1 does not substantially increase the flow rate due to the proximity of the bottom wall (membrane), where the velocity is zero. In addition, only one neighboring row is partially affected by the change in flow field, compared to two if the vacancies were in a middle row. Nevertheless, close to the membrane the potential difference $\Delta \phi$ is higher than further away (positions with same $x$ and larger $y$, see also Fig.~\ref{fig:example}(f)). Since the reaction rate is therefore higher, the shadowing effect will also be more pronounced in lower rows (see also Fig.~\ref{fig:example}(e)). Therefore, having vacancies in a lower row is beneficial for reducing the concentration overpotential. This results in a maximum of $j$ when vacancies are in row 2, where the boundary effect from the wall on the flow speed is already negligible. The design with vacancies in row 7 is the worst since it suffers from the confinement effect and there is little gain in reducing shadowing since $\Delta \phi$ is smaller than in other rows. We also investigated the case where we remove every third cylinder instead of every second one (dashed lines in Fig.~\ref{fig:vacrows}) and we observe the same qualitative trend of $j$ as a function of the row number, implying that indeed the main reason of the non-monotonic behavior is the distance from the walls, or in other words the top--bottom symmetry breaking, rather than trailing effects along the flow direction. For the lattice spacing of our study, when there are vacancies for every second cylinder the current extracted is larger than when the vacancies are for every third cylinder, since shadowing effects are smaller. 

\begin{figure}
\includegraphics[width=0.49\textwidth]{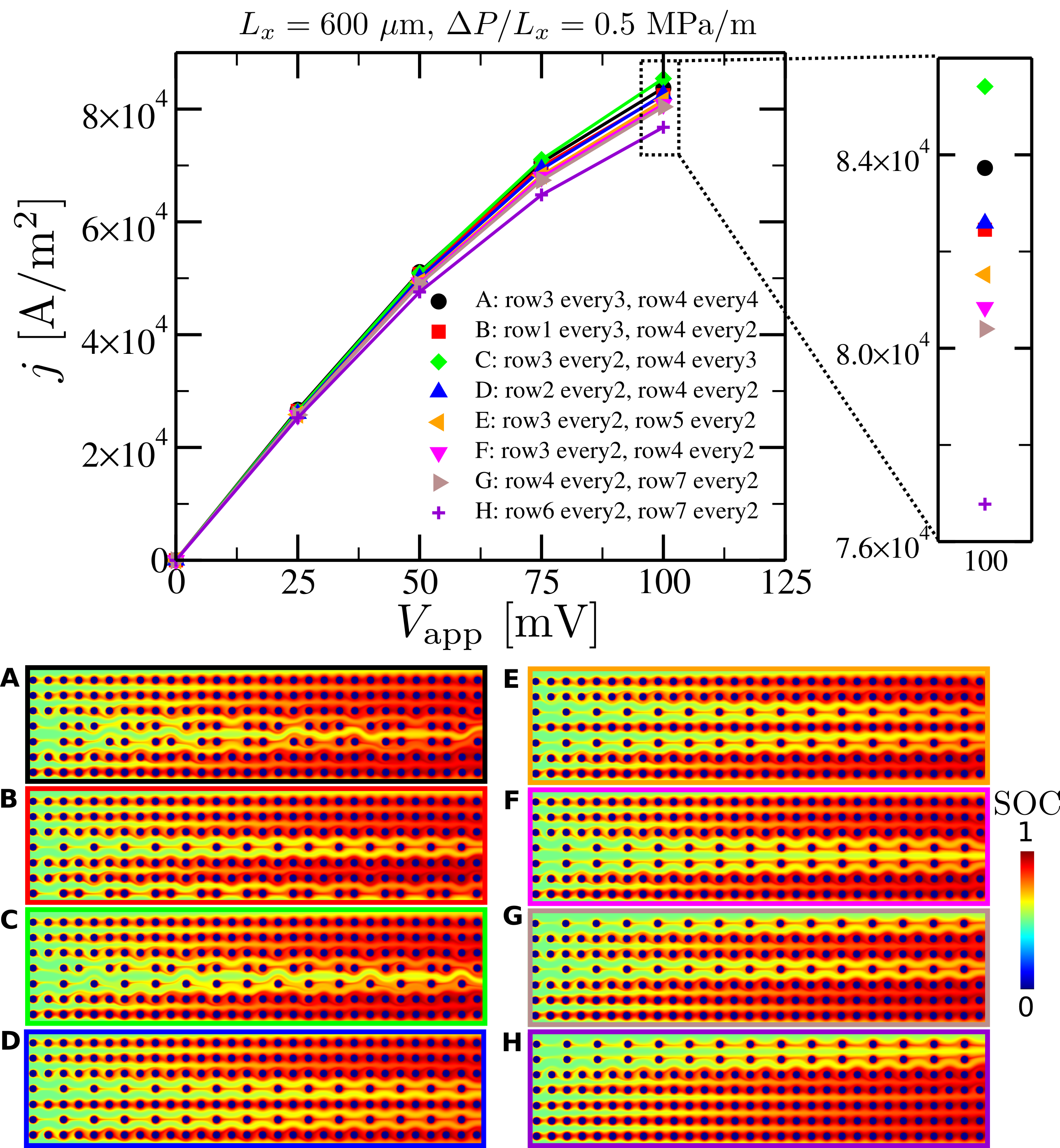}
\caption{Current density as a function of applied voltage for several configurations featuring two rows with vacancies with location and frequency as indicated in the legend. Corresponding snapshots of the state of charge for $V_{\text{app}}=\SI{100}{\mV}$ are also shown.}
\label{fig:vacvar1}
\end{figure}

Next, we investigate several combinations of pair of rows with vacancies. We report in Fig.~\ref{fig:vacvar1} the most relevant cases. Note that not all the configurations have the same total reactive surface area, since they have rows with vacancies with different frequency. In particular, configuration a has the largest surface area because of a smaller number of vacancies, whereas configurations d to h have the smallest surface area. Again, we note that the total surface area alone is not a direct indicator of the performance, as it can be observed from the polarization curves (especially from the inset for $V_{\text{app}}=\SI{100}{\mV}$). Focusing first on the configurations with same surface area, we observe the following performance ranking: D, E, F, G, H. Lowest $j$ can be extracted from configuration H because the vacancies are in the top rows (number 6 and 7), where $\Delta \phi$ is low; they are next to a wall, where little increase of the fluid speed can be achieved; they are next to each other and the concentration profile of only one adjacent row (number 5) is perturbed, so shadowing effects are not efficiently mitigated. Selecting two non-adjacent top rows (configuration G), two adjacent middle rows (F), two non-adjacent middle rows (E), are all improvements over the previous design. Finally, configuration D exhibits even higher current since the vacancies are placed in row 2, that is the optimal row for this geometry and operating conditions, as illustrated above. When allowing for different frequency of the vacancies, the number of combinations clearly increases and similar arguments on favorable rows can be made. In addition, we now observe that having two rows next to each other with a different vacancies frequency (e.g.\@ configurations A and C) can boost the performance. In fact, the trails are efficiently redirected and there is larger opportunity for species mixing and replenishing the reactant concentration, i.e.\@ minimizing shadowing effects. In this example, configuration C has the largest $j$ despite having less reactive surface area than configuration A.

\begin{figure}
\includegraphics[width=0.49\textwidth]{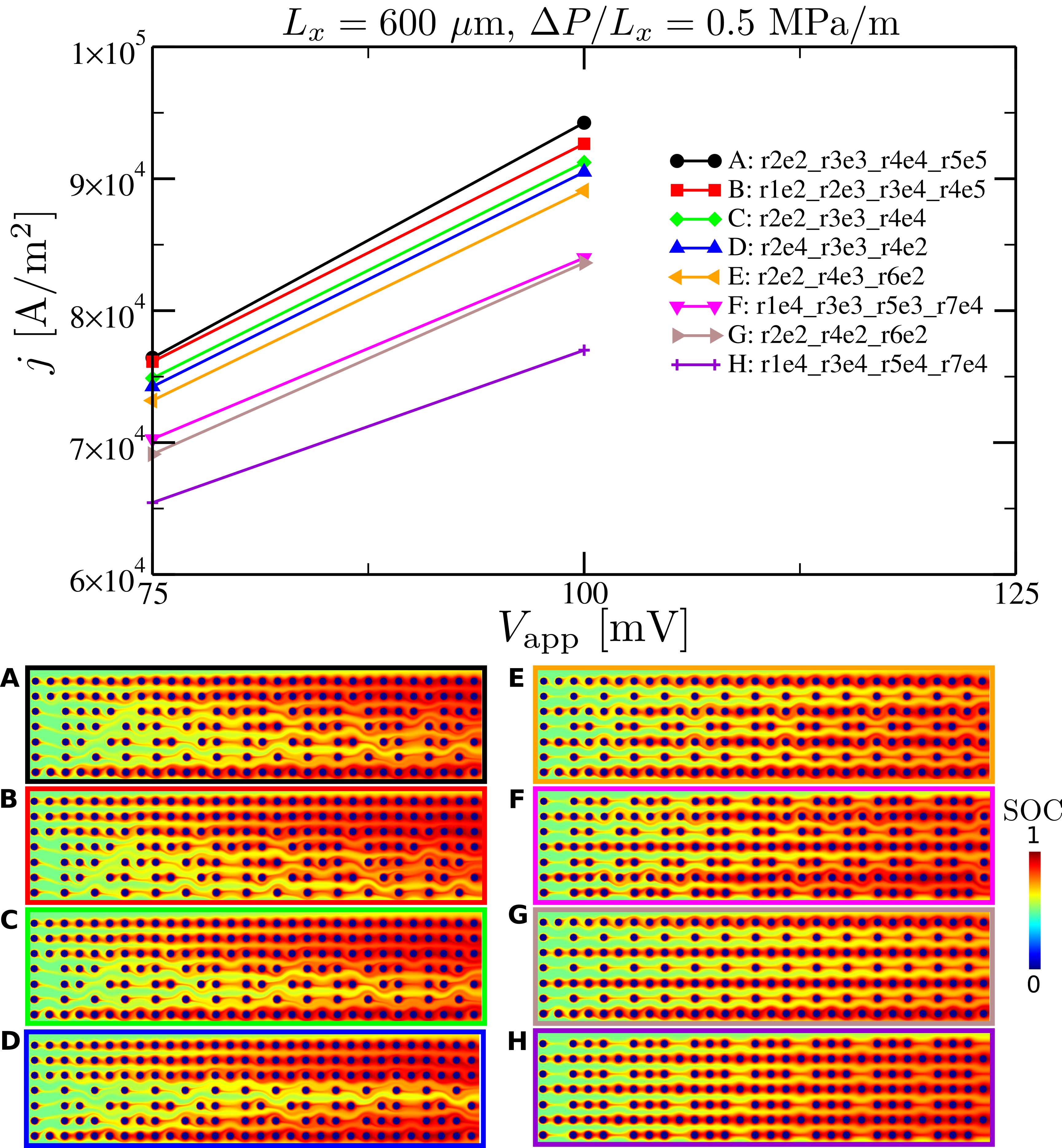}
\caption{Polarization curves for configurations featuring three or four rows with vacancies with location ($\text{r}=\text{row}$) and frequency ($\text{e}=\text{every}$) as indicated in the legend. Corresponding snapshots of the state of charge for $V_{\text{app}}=\SI{100}{\mV}$ are shown.}
\label{fig:vacvar2}
\end{figure}

To further gain insight in how to engineer the location and frequency of the vacancies to increase electrode performance, in Fig.~\ref{fig:vacvar2} we report selected examples of lattices with three or four rows of vacancies. The investigated configurations have different number of vacancies,
and therefore different porosity and surface area. They are ordered according to the extracted current density $j$. At the bottom, we find configurations (G, H) with vacancies in non-adjacent rows and with same frequency. Configuration H performs significantly worse since there are not enough vacancies and some are placed just next to the walls. Configuration F is a small improvement compared to G thanks to a mismatch in vacancies frequency, even though they are placed in non-adjacent rows and some next to the wall. Configuration E has fewer rows with vacancies, that are however more frequent and in middle rows. The top four configurations (A, B, C, D) consist of adjacent rows with vacancies at different frequencies, so significant mixing occurs. In addition, we can now have density gradients from the bottom (membrane) to the top by adjusting the frequency of vacancies. When looking at configuration C and D, the only difference is that the former consists of an increasing density gradient (more vacancies at the bottom and less at the top) and vice versa for the latter. The comparison indicates that better performance can be obtained with a bottom-to-top decreasing density gradient. Adding an additional row of vacancies allows to extend the gradient within the microstructure. However, the exact choice of rows with vacancies has still an effect of the microstructure performance. Indeed, starting the gradient from row 1 (configuration B) leads to a smaller improvement than when starting from row 2 (configuration A), consistent with the previous observations indicating that vacancies in row 1 are less effective than ones in row 2.

\subsection{Disordered configurations}
\begin{figure}
\includegraphics[width=0.49\textwidth]{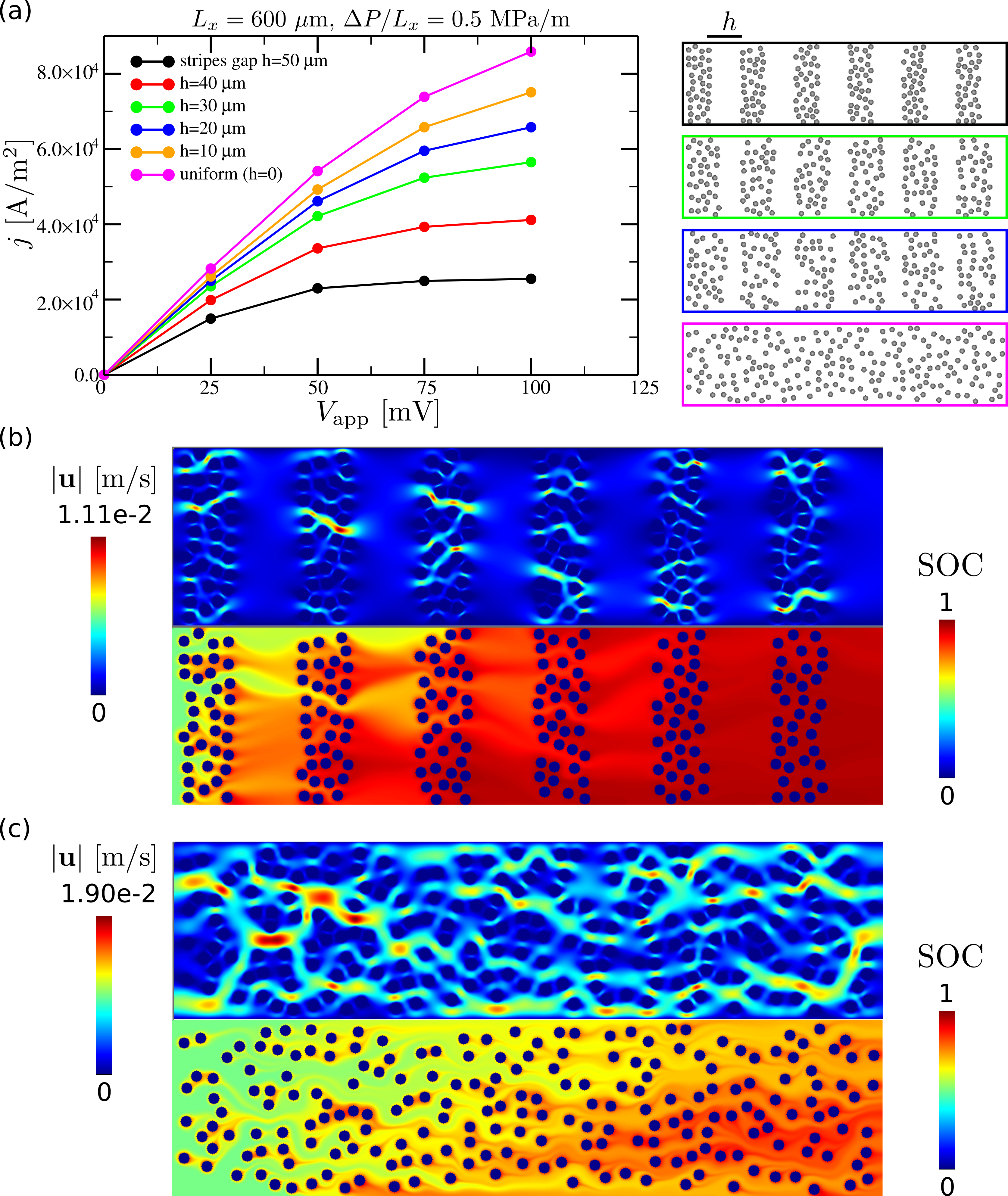}
\caption{(a) Polarization curves and schematics of configurations with alternating large pores and disordered arrangements of cylinders. Snapshots of velocity and state of charge for configurations with (b) $h=\SI{50}{\um}$ and (c) $h=\SI{0}{\um}$ (uniformly distributed cylinders) obtained for $V_{\text{app}}=\SI{50}{\mV}$. Uniformly random configurations outperform striped configurations featuring two length-scales.}
\label{fig:vstrip}
\end{figure}

We have seen that battery performance can be increased by moving away from the regular fully ordered lattice and introducing vacancies in the system, designed in a way such that the mixing of the reactive species is enhanced. Generally, mixing is intrinsically present in disordered configurations, therefore raising the question whether they could naturally represent a more suitable electrode microstructure. Additionally, experimental studies reported that structures featuring two length scales, particularly carbon cloths, are very performant.~\cite{Forner-Cuenca2019a,Tenny2020} Intuitively, this can be attributed to the fact that the larger voids allow for replenishing of the reactant concentration, in the same spirit of our observations for configurations with vacancies in a single row, where product trails have time to diffuse away before investing the downstream material. In this section, we consider examples of disordered configurations, starting from uniformly randomly distributed cylinders and progressively separating them in alternating vertical stripes of disordered material and empty space of size $h$ (see Fig.~\ref{fig:vstrip}(a)). In all cases, we consider 210 cylinders, therefore we have the same total porosity and surface area as for the reference ordered lattice configuration. The polarization curves obtained at same pressure-driven flow clearly reveal a smooth increase in performance from well separated to uniform disordered arrangements. One reason for this is the lower permeability of the striped configurations compared to the uniform arrangements. In fact, overall and local velocities are significantly lower in the two-scale configurations (see panels b and c of Fig.~\ref{fig:vstrip}), due to the presence of local regions with much higher material density. In addition, despite the fact that within the stripe the cylinders are uniformly randomly distributed and therefore should favor mixing, they are tightly packed and the space in between the stripes is not sufficient to replenish the reactant concentration. We can conclude then that uniformly disordered configurations can be better than some two length-scale structures. At present, it is however difficult to generalize this statement to different structures with two length scales, including carbon cloths where 3D effects might also play a dominant role, or for different porosities due to the subtle interplay of the typical length scales associated to geometrical features and physical processes, such as diffusion.

\subsection{Engineered vacancies can be better than disordered configurations}
\begin{figure}
\includegraphics[width=0.49\textwidth]{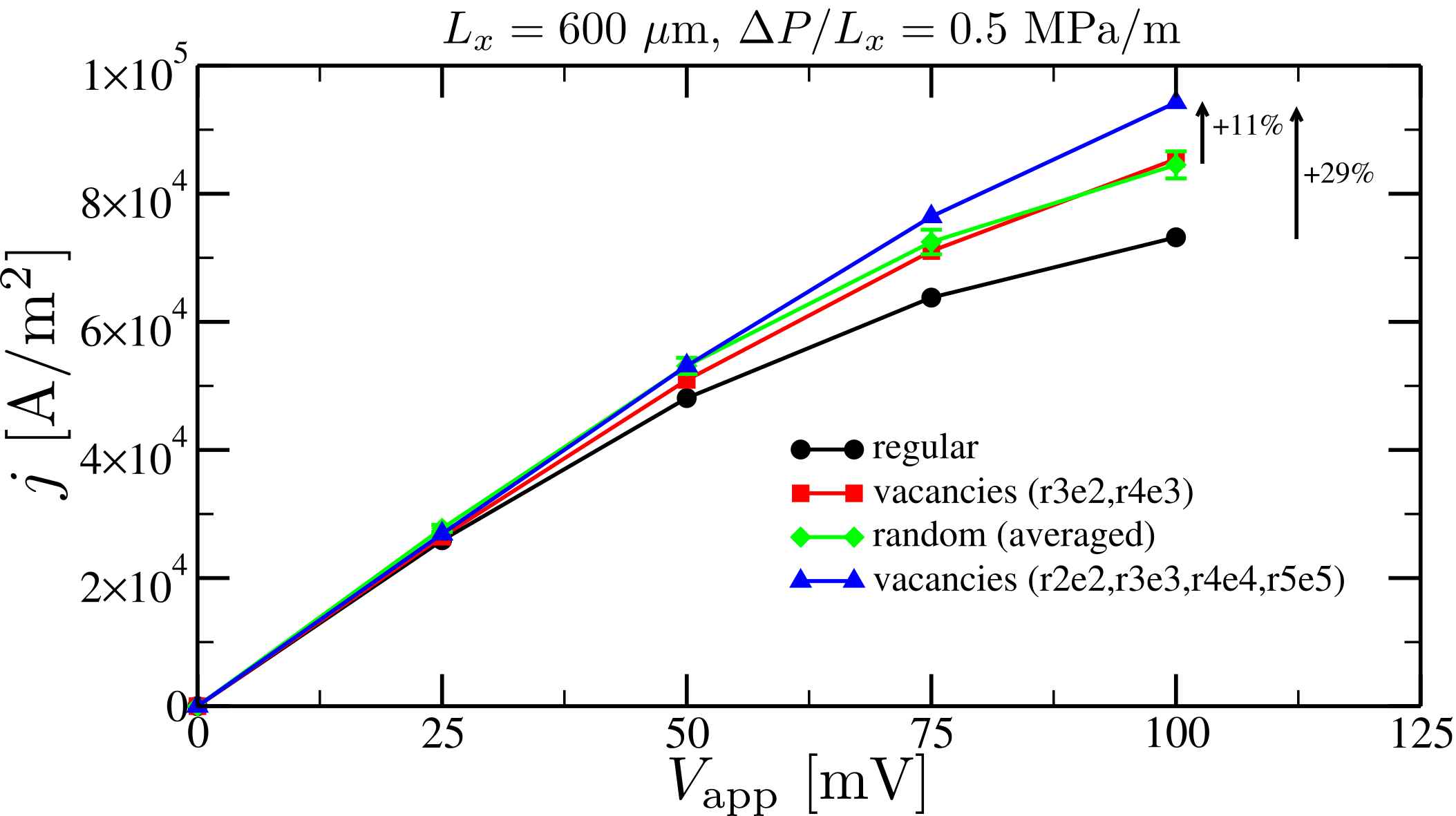}
\caption{Comparison between the polarization curves of the key configurations investigated in this study. The configuration featuring four rows with vacancies arranged in a density gradient fashion outperforms both the regular lattice and the disordered (random) structure. Data for the random structure are averaged over several equivalent realizations.}
\label{fig:disvac}
\end{figure}

Finally, in Fig.~\ref{fig:disvac} we compare the polarization curves of the key configurations investigated in our study. Disordered configurations show better performance than the fully ordered regular configuration, due to the fact that shadowing effects that are characteristic of the latter are intrinsically avoided in the former, as detailed in the previous section. Nevertheless, it is also apparent that a significant performance boost can be obtained by carefully engineering the location of the vacancies, allowing for a better dispersion of the reactive species. The 10\% gain observed for this system size ($L_x=\SI{600}{\um}$) and operating conditions is likely to become even more significant for longer systems (following the same considerations associated to Fig.~\ref{fig:syssize}) and generally when mass transport limitations are more severe (i.e.\@ even larger applied potential and/or lower reactant concentration). This evidence suggests that by further exploring the design space of microstructures and systematically pinpointing which structural features cause an increase or decrease in RFB performance, it will be possible to identify novel designs that outperform the current commercial electrodes that are all feature disordered structures.

\section{\label{sec:con}Conclusion}
In summary, we introduced and showcased a simulation method for capturing flow, mass and charge transport in RFBs based on a parallel computational framework (AMReX+HYPRE). Our systematic investigation on 2D microstructures identified how regular lattices can be modified to achieve higher electrode performance. Our step back in complexity allowed a step forward in the understanding of the non-trivial processes associated to the electrode microstructure and identified guidelines on how to introduce vacancies and gradients to allow for better species mixing, reducing shadowing effects and counter potential gradients. Future work should be dedicated to understand how these concepts translate to more complex 3D structures. 

We want to remark that several extensions have already been implemented within the AMReX framework, and they can be explored in future works on RFBs. These include simulations of 3D structures, use of (adaptive) mesh refinement, and different algorithms within the context of the EB method (e.g.\@ a Godunov scheme for handling the convective terms). Future studies could also investigate the most efficient way of updating concentrations and potential, e.g.\@ using Newton's method to simultaneously update both quantities and possibly obtaining a faster convergence within a single time-step, at the cost of solving a much larger system of linear equations. 

Therefore, the proposed framework holds the potential to efficiently run massively parallel simulations, with mesh refinement to decrease computational costs, opening up to the possibility of performing simulation-based design optimization for specific electrochemical flow systems. Here, we used physical parameters of AQDS-based RFB, representing therefore a kinetically facile system. We also did not consider any fiber-level mass transfer coefficient, so we neglected the possible influence of e.g. surface roughness. Fine-tuning of the model and quantitative comparison with experimental data can be achieved once realistic 3D structures are simulated. Given the exciting progress on visualization of the flow and the electrochemical processes in RFBs~\cite{Bazylak2009, Tariq2018, Wong2021}, the combination of direct numerical simulations and experimental results will be able to shine new light into the role of electrode microstructure on battery performance.

\begin{acknowledgments}
We thank Michael Aziz, Meisam Bahari, and Kiana Amini for useful discussions on redox flow batteries. We thank Michael Aziz and Michael Emanuel for critical feedback on the manuscript. The authors gratefully acknowledge the funding provided by DOE BES (award number DE-SC0020170). This research used resources of the National Energy Research Scientific Computing Center (NERSC), a U.S.\@ Department of Energy Office of Science User Facility located at Lawrence Berkeley National Laboratory.
\end{acknowledgments}

\section*{Data Availability Statement}

The data that support the findings of this study are available upon reasonable request.


%

\end{document}